\documentclass[twocolumn]{aastex63}

\graphicspath{{./}{figures/}}

\received{--}
\revised{--}
\accepted{--}
\submitjournal{ApJL}

\usepackage{multirow}
\usepackage{txfonts}
\usepackage{graphicx}
\usepackage[outdir=./]{epstopdf}
\usepackage{natbib}
\usepackage{xcolor, color}
\usepackage{booktabs}
\usepackage{threeparttable}
\usepackage{hyperref}
\hypersetup{linkcolor=red,citecolor=blue,filecolor=cyan,urlcolor=magenta}
\usepackage{comment}

\DeclareUnicodeCharacter{2212}{-}
\newcommand{\kms}{km s$^{-1}$}

\newcommand{\meth}{CH$_3$OH}

\shortauthors{De Simone et al.}
\begin{document}

\title{Tracking the ice mantle history in the Solar-type Protostars of NGC 1333 IRAS 4}

\correspondingauthor{Marta De Simone}
\email{marta.desimone@univ-grenoble-alpes.fr}

\author[0000-0001-5659-0140]{Marta De Simone}
\affiliation{Univ. Grenoble Alpes, CNRS, IPAG, 38000 Grenoble, France}
\affiliation{INAF, Osservatorio Astrofisico di Arcetri, Largo E. Fermi 5, 50125 Firenze, Italy}

\author[0000-0001-9664-6292]{Cecilia Ceccarelli}
\affiliation{Univ. Grenoble Alpes, CNRS, IPAG, 38000 Grenoble, France}

\author[0000-0003-1514-3074]{Claudio Codella}
\affiliation{INAF, Osservatorio Astrofisico di Arcetri, Largo E. Fermi 5, 50125 Firenze, Italy}
\affiliation{Univ. Grenoble Alpes, CNRS, IPAG, 38000 Grenoble, France}

\author[0000-0002-8502-6431]{Brian E. Svoboda}
\affiliation{National Radio Astronomy Observatory\footnote{\label{fn:NRAO}The National Radio Astronomy Observatory is a facility of the National Science Foundation operated under cooperative agreement by Associated Universities, Inc.}, 1003 Lopezville Rd, Socorro, NM 87801, USA}

\author[0000-0002-7570-5596]{Claire J. Chandler}
\affiliation{National Radio Astronomy Observatory\footnote{\label{fn:NRAO}The National Radio Astronomy Observatory is a facility of the National Science Foundation operated under cooperative agreement by Associated Universities, Inc.}, 1003 Lopezville Rd, Socorro, NM 87801, USA}

\author[0000-0003-0167-0746]{Mathilde Bouvier}
\affiliation{Univ. Grenoble Alpes, CNRS, IPAG, 38000 Grenoble, France}

\author[0000-0002-9865-0970]{Satoshi Yamamoto}
\affiliation{Department of Physics, The University of Tokyo, Bunkyo-ku, Tokyo 113-0033, Japan}

\author[0000-0002-3297-4497]{Nami Sakai}
\affiliation{The Institute of Physical and Chemical Research (RIKEN), 2-1, Hirosawa, Wako-shi, Saitama 351-0198, Japan}

\author[0000-0001-8227-2816]{Yao-Lun Yang}
\affiliation{The Institute of Physical and Chemical Research (RIKEN), 2-1, Hirosawa, Wako-shi, Saitama 351-0198, Japan}
\affiliation{Department of Astronomy, University of Virginia, Charlottesville, VA 22904, USA}

\author[0000-0003-1481-7911]{Paola Caselli}
\affiliation{Max-Planck-Institut f\"ur extraterrestrische Physik (MPE), Giessenbachstrasse 1, 85748 Garching, Germany}

%\author[0000-0002-5789-6931]{Cecile Favre}
%\affiliation{Univ. Grenoble Alpes, CNRS, IPAG, 38000 Grenoble, France}

\author[0000-0002-9397-3826]{Bertrand Lefloch}
\affiliation{Univ. Grenoble Alpes, CNRS, IPAG, 38000 Grenoble, France}

\author[0000-0003-2300-2626]{Hauyu Baobab Liu}
\affiliation{Academia Sinica Institute of Astronomy and Astrophysics (ASIAA), No. 1, Section 4, Roosevelt Road, Taipei 10617, Taiwan}

\author[0000-0002-6729-3640]{Ana L\'opez-Sepulcre}
\affiliation{Institut de Radioastronomie Millim\'etrique (IRAM), 300 rue de la Piscine, 38400 Saint-Martin d'H\`eres, France}
\affiliation{Univ. Grenoble Alpes, CNRS, IPAG, 38000 Grenoble, France}

\author[0000-0002-5635-3345]{Laurent Loinard}
\affiliation{Instituto de Radioastronomía y Astrofísica, Universidad Nacional Autónoma de México Apartado 58090, Morelia, Michoacán, Mexico}

\author[0000-0002-3972-1978]{Jaime E. Pineda}
\affiliation{Max-Planck-Institut f\"ur extraterrestrische Physik (MPE), Giessenbachstrasse 1, 85748 Garching, Germany}

\author[0000-0003-1859-3070]{Leonardo Testi}
\affiliation{ESO, Karl Schwarzchild Srt. 2, 85478 Garching bei M\"unchen, Germany
}
\affiliation{INAF, Osservatorio Astrofisico di Arcetri, Largo E. Fermi 5, 50125 Firenze, Italy}
\affiliation{Excellence Cluster Origins, Boltzmannstrasse 2, D-85748 Garching bei München, Germany}

\begin{abstract}

% the AAS Journals, the Astrophysical Journal (ApJ), the Astrophysical Journal Letters (ApJL), and Astronomical Journal (AJ), all have a 250 word limit for the abstract
To understand the origin of the diversity observed in exoplanetary systems, it is crucial to characterize the early stages of their formation, represented by Solar-type protostars. 
Likely, the gaseous chemical content of these objects directly depends on the composition of the dust grain mantles formed before the collapse. 
Directly retrieving the ice mantle composition is challenging, but it can be done indirectly by observing the major components,  {such as NH$_3$ and CH$_3$OH at cm wavelengths}, once they are released into the gas-phase during the warm protostellar stage. 
We observed several CH$_3$OH and NH$_3$ lines toward three Class 0 protostars in NGC1333 (IRAS 4A1, IRAS 4A2, and IRAS 4B), at  {high angular resolution} (1$''$; $\sim$300 au) with the VLA interferometer  {at 24-26 GHz}. 
Using a non-LTE LVG analysis, we derived a similar NH$_3$/CH$_3$OH abundance ratio in the three protostars ($\leq$ 0.5, 0.015--0.5, and 0.003-0.3 for IRAS 4A1, 4A2, and 4B, respectively). Hence, we infer they were born from pre-collapse material with similar physical conditions.
Comparing the observed abundance ratios with astrochemical model predictions, we constrained the dust temperature at the time of the mantle formation to be $\sim$17 K, which coincides with the average temperature of the southern NGC 1333 diffuse cloud. 
We suggest that a brutal event started the collapse that eventually formed IRAS 4A1, 4A2 and 4B, which, therefore, did not experience the usual pre-stellar core phase.
This event could be the clash of a bubble with NGC 1333 south, that has previously been evoked in the literature.
\end{abstract}


\section{Introduction} \label{sec:intro}
The thousands of exoplanets discovered so far (e.g., \url{http://exoplanet.eu/}) provide clear evidence of the incredible variety of planetary systems, different from each other and from our Solar System.
To understand the origin of such diversity, it is crucial to characterize the early stages of the formation of a planetary system.
To study the diversity of these early stages, a powerful observational diagnostic tool is their chemical composition \citep{ceccarelli_extreme_2007,sakai_warm_2013}. 
Indeed, the chemical complexity in star-forming regions starts at the very beginning of the process, during the pre-collapse phase. At this stage, icy mantles form on interstellar grains and grow rich in hydrogenated species. Then, during the warm ($>$100 K) protostellar phase, the ice mantle species are released in the gas-phase through ice mantle sublimation  {\citep{viti_evaporation_2004, herbst_complex_2009, caselli_our_2012, oberg_astrochemistry_2021}}.
Therefore, the composition of the icy mantles is crucial in establishing the available gaseous chemical content that could explain the observed chemical diversity. 

Infrared (IR) absorption observations towards Solar-type protostars have shown that the icy mantles major components are H$_2$O, CO, CO$_2$, CH$_4$, NH$_3$, and CH$_3$OH \citep[e.g.,][]{boogert_observations_2015}. 
However, these observations can only be obtained toward sources with enough bright IR continuum emission, making it very difficult to characterize the ice mantles of deeply embedded protostars, and even more of prestellar cores. 
Another possibility to study the ice mantle chemical composition is to observe their major components once they are released into the gas-phase during the warm protostellar phase  {\citep[see e.g.,][]{whittet_observational_2011}}.

In this context, NH$_3$ and CH$_3$OH are the best, if not the only, major components of the icy mantles that  {trace the hot central protostar with ground-based high angular resolution observations}. Indeed,  {CO is confused with the surrounding cloud}, CO$_2$ and CH$_4$ do not have dipole moments, and H$_2$O is hampered by the terrestrial atmosphere.
% {Note that, NH$_3$ and CH$_3$OH high excitation transitions permit to isolate the emission from only the hot central core. }
This indirect method relies on the knowledge of the species formation pathways and that they are efficiently released into the gas with the water ice mantle

 {This indirect method relies on the knowledge of the species formation pathways and that they are efficiently released into the gas with the water ice mantle. In this case,} the formation paths of NH$_3$ and CH$_3$OH are very well studied: they are mainly formed in the prestellar phase on the icy grain mantles through N and CO hydrogenation \citep{watanabe_efficient_2002,rimola_combined_2014, song_tunneling_2017,jonusas2020} even if, for NH$_3$, a gas-phase contribution with a subsequent depletion cannot be excluded \citep[e.g.,][see also Figure \ref{fig:scheme_nh3}]{le_gal_interstellar_2014,pineda_interferometric_2022,caselli_central_2022}. 
 {In the following, we assume that the observed composition of the gas phase reflects that of the icy mantles, since the latter, and all the species within, sublimate during the hot corino phase when the dust temperature reaches the water sublimation temperature ($\geq$100 K).}
Once released in the gas-phase, both methanol and ammonia undergo chemical reactions that can alter their abundances. However, the continuous infall of newly-sublimated material ensures that the gas-phase abundance of these two species indeed reflects the one on the grain mantles.

Finally, the NH$_3$/CH$_3$OH abundance ratio on the grain mantles depends only on the physical conditions of the material before the collapse, namely  {gas density, dust temperature and ice mantle formation timescale. 
These parameters regulate, for example, the N and CO hydrogenation efficiency and their residence time on the mantles} \citep[e.g.,][]{caselli_chemical_1993,taquet_multilayer_2012, aikawa_chemical_2020}. 
These physical conditions can be different based on the dynamical history of the single object, which could be affected by external factors (e.g., cloud-cloud collisions, supernovae explosions, etc.). 
%Moreover, both molecules are key participants in the formation of more complex N- and O-bearing species.

In this Letter, we investigate the icy mantle composition, through the relative abundance of NH$_3$ and CH$_3$OH, of three Class 0 protostars. %: IRAS 4A1, IRAS 4A2 and IRAS 4B. 
We used the VLA interferometer to trace the inner  {300 au at centimeter} wavelengths, where the dust is more likely optically thin \citep[see e.g.,][]{li_systematic_2017, ko_resolving_2020, de_simone_hot_2020}.
 {We also used previous millimeter observations of methanol and its isotopologues  \citep{taquet_constraining_2015,yang_perseus_2021} to estimate the dust absorption contribution, and to constrain the CH$_3$OH column density in case of optically thick emission.}

The three targeted sources are located in the southern filament of the Perseus/NGC1333 region \citep[$\sim300$ pc;][]{zucker_mapping_2018}: the protobinary system IRAS 4A, composed of IRAS 4A1 and IRAS 4A2 (hereafter 4A1 and 4A2) separated by 1.8$''$ ($\sim$540 au), and IRAS 4B (hereafter 4B) located $\sim30''$ south-east of 4A2.
The three sources are known to be hot corinos\footnote{Hot corinos are compact ($<$100 au), hot ($\geq$100 K) and dense (n$_{\rm H_2}\geq$10$^7$ cm$^{-3}$) regions \citep[e.g.,][]{ceccarelli_hot_2004} around Solar-type protostars, enriched in interstellar Complex Organic Molecules \citep[iCOMs;][]{herbst_complex_2009,ceccarelli_seeds_2017}.}
\citep{sakai_detection_2006,bottinelli_hot_2007,taquet_constraining_2015,lopez-sepulcre_complex_2017,de_simone_glycolaldehyde_2017,de_simone_hot_2020}.
 {Recently, observational evidence pointed out that this filament, where the three protostars lie, could have been shaped by the clash of an expanding bubble with NGC1333 \citep{dhabal_connecting_2019, de_simone_bubble_2021}. In particular, \citet{dhabal_connecting_2019} suggested that this clash could have been responsible for the formation of the protostars.}

\section{Observations and results} \label{sec:obs}
%The data used for this work are the 
We used VLA observations in K-band (project ID: 18B-166) described in \citet{de_simone_hot_2020}. 
In summary, we targeted ten CH$_3$OH and five NH$_3$ lines, with frequencies from 23.8 to 26.4 GHz and a large range of upper level energies (E$_{\rm up}$), %from 36 to 175 K for CH$_3$OH and from 24 to 640 K for NH$_3$ 
(Table \ref{tab:spectral_params&fit_res}). They were associated with 13 spectral windows with $\sim$0.017 MHz ($\sim$0.2 \kms) spectral channels and 1$''$ angular resolution. The absolute flux calibration error is $\leq$15\%\footnote{https://science.nrao.edu/facilities/vla/docs/manuals/oss/performance/fdscale}.
Data reduction and cleaning process were performed using CASA\footnote{https://www.casa.nrao.edu/}, and data analysis and images using GILDAS\footnote{http://www.iram.fr/IRAMFR/GILDAS}.
%Further details can be found in \citet{de_simone_hot_2020}. 
The continuum is obtained by averaging line-free channels from all the spectral windows. We self-calibrated, in phase and amplitude, using the line-free continuum channels and applied the solutions to both the continuum and molecular lines \citep[see][]{de_simone_hot_2020}.
The continuum subtracted cubes, were smoothed to 1 \kms. They were cleaned using a manually corrected threshold mask for each channel, and a multiscale deconvolution (scales = [0, 5, 15, 18, 25]) with natural weighting. The synthesized beams are in Table \ref{tab:spectral_params&fit_res}. 
%The half power primary beam is $\sim 110''(?)$.

%\section{Results}
\begin{table*}
 \centering
    \caption{Spectral parameters, synthesized beams and the Gaussian and Hyperfine fit results for CH$_3$OH and NH$_3$, respectively, extracted toward the protostars continuum peak.
    }\label{tab:spectral_params&fit_res}
    \hspace{-2cm}
    \begin{tabular}{l|cccc|cccc}
        \hline
        \hline
        \multirow{2}{*}{Transition} & \multirow{2}{*}{Frequency$^{(a)}$} & \multirow{2}{*}{E$_{\rm up}^{(a)}$}& \multirow{2}{*}{logA$_{\rm ij}^{(a)}$} & Synthesized Beam & \multicolumn{4}{c}{Source} \\
        %& \multicolumn{4}{c|}{IRAS 4A2} & \multicolumn{4}{c}{IRAS 4A1} \\
            &   &   &   & maj $\times$ min (PA)  &   $\rm \int T_BdV$ & V$_{\rm peak}^{b}$ & FWHM$^{b}$ & RMS  \\
            %&   $\rm \int T_BdV^d$ & V$_{\rm peak}^{b}$ & FWHM$^{b}$ & RMS %&   $\rm \int T_BdV^d$ & V$_{\rm peak}^{b}$ & FWHM$^{b}$ & RMS \\
            & [GHz] & [K] &  & [$''\times ''$ ($^\circ$)] & [K km s$^{-1}$] & [km s$^{-1}$] & [km s$^{-1}$] & [K]  \\
            %& [K km s$^{-1}$] & [km s$^{-1}$] & [km s$^{-1}$] & [K] & [K km s$^{-1}$] & [km s$^{-1}$] & [km s$^{-1}$] & [K] \\
        \hline
         \multicolumn{5}{c|}{\multirow{2}{*}{CH$_3$OH}} &  \multicolumn{4}{c}{IRAS 4B} \\
         \multicolumn{5}{c|}{} &  \multicolumn{4}{c}{$\rm 03^h29^m12\fs02$, $\rm 31^\circ13'07\farcs9$} \\
        \hline
        3(2,1)-3(1,2) E & 24.9287 & 36 & -7.2 & %$1.04\times1.01 \ (-11)$  %continuum beam
        $0.97\times0.95 \ (-12)$ & 17(2) & +6.9(0.1) & 1.5(0.2) & 1.2 \\
        4(2,2)-4(1,3) E & 24.9334 & 45 & -7.1 & $0.97\times0.95 \ (-12)$ & 21(3) & +6.9(0.2) & 2.2(0.4) & 1.2 \\
        2(2,0)-2(1,1) E & 24.9343 & 29 & -7.2 & $0.97\times0.95 \ (-12)$ & 17(3) & +6.9(0.2) & 2.3(0.4) & 1.2 \\
        5(2,3)-5(1,4) E & 24.9590 & 57 & -7.1 & $0.97\times0.95 \ (-12)$ & 17(2) & +6.8(0.1) & 1.5(0.2) & 1.2   \\
        6(2,4)-6(1,5) E & 25.0181 & 71 & -7.1 & $0.97\times0.95 \ (-19)$ & 21(3) & +6.9(0.2) & 2.3(0.5) & 1.4 \\
        7(2,5)-7(1,6) E & 25.1248 & 87 & -7.1 & $0.98\times0.95 \ (-21)$ & 19(2) & +6.8(0.1) & 1.7(0.2) & 1.5 \\
        8(2,6)-8(1,7) E & 25.2944 & 106 & -7.0 & $0.96\times0.94 \ (-11)$ & 19(2) & +6.9(0.1) & 1.9(0.3) & 1.2 \\
        9(2,7)-9(1,8) E & 25.5414 & 127 & -7.0 & $0.96\times0.92 \ (-50)$ & 22(3) & +6.8(0.1) & 1.8(0.3) & 1.4 \\
        10(2,8)-10(1,9) E & 25.8782 & 150 & -7.0 & $0.97\times0.93 \ (-35)$ & 16(2) & +7.1(0.1) & 1.9(0.3) & 1.1 \\
        11(2,9)-11(1,10) E & 26.3131 & 175 & -6.9 & $0.94\times0.91 \ (-35)$ & 14(2) & +6.6(0.2) & 1.4(0.5) & 1.4 \\
        \hline
        \multicolumn{5}{c|}{\multirow{2}{*}{NH$_3$}} &  \multicolumn{4}{c}{IRAS 4B} \\
         \multicolumn{5}{c|}{} &  \multicolumn{4}{c}{$\rm 03^h29^m12\fs02$, $\rm 31^\circ13'07\farcs9$} \\
        \hline
        (3,3) & 23.8701 & 124 & -6.6 & $1.00\times0.95 \ (+6)$ & 64(8) & +7.2(0.2) & 1.7(0.5) & 1.5  \\
        %& 169(20) & 6.9(0.1) & 2.3(0.1) & 1.3 & 121(16) & 6.4(0.1) & 2.7(0.2) & 1.3 \\
        (4,4) & 24.1394 & 201 & -6.5 & $0.99\times0.94 \ (-2)$ & 41(7) & +7.1(0.1) & 1.7(0.1) & 1.4  \\
        %& 89(14) & 6.9(0.1) & 1.8(0.1) & 1.2 & 90(13) & 6.5(0.1) & 1.9(0.1) & 1.0 \\
        (5,5) & 24.5329 & 296 & -6.5 & $0.99\times0.95 \ (-3)$ & 39(7) & +6.6(0.4) & 2.5(0.7) & 1.4   \\
        %& 54(14) & 6.9(0.1) & 1.7(0.2) & 1.2 & 87(12) & 6.4(0.1) & 2.5(001 & 1.0 \\
        (6,6) & 25.0560 & 409 & -6.5 & $0.97\times0.95 \ (-18)$ & 37(7) & +6.1(0.2) & 1.7(0.4) & 1.4   \\
        %& 62(11) & 6.5(0.1) & 2.3(0.1) & 1.1 & 87(11) & 6.7(0.1) & 2.9(0.2) & 1.1 \\
        (7,7) & 25.7151 & 639 & -6.4 & $0.96\times0.92 \ (-45)$ & 27(7) & +6.9(0.3) & 1.7(0.1) & 1.4  \\
        %& 40(8) & 7.1(0.1) & 1.8(0.1) & 1.0 & 46(9) & 6.4(0.1) & 1.9(0.2) & 1.1 \\
        \hline
         \multicolumn{5}{c|}{\multirow{2}{*}{NH$_3$}} &  \multicolumn{4}{c}{IRAS 4A2} \\
         \multicolumn{5}{c|}{} &  \multicolumn{4}{c}{$\rm 03^h29^m10\fs43$, $\rm 31^\circ13'32\farcs1$} \\
        \hline
        (3,3) & 23.8701 & 124 & -6.6 & $1.00\times0.95 \ (+6)$ & 169(20) & +6.9(0.1) & 2.3(0.1) & 1.3 \\
        (4,4) & 24.1394 & 201 & -6.5 & $0.99\times0.94 \ (-2)$ &  89(14) & +6.9(0.1) & 1.8(0.1) & 1.2 \\
        (5,5) & 24.5329 & 296 & -6.5 & $0.99\times0.95 \ (-3)$ & 54(14) & +6.9(0.1) & 1.7(0.2) & 1.2 \\
        (6,6) & 25.0560 & 409 & -6.5 & $0.97\times0.95 \ (-18)$ &  62(11) & +6.5(0.1) & 2.3(0.1) & 1.1  \\
        (7,7) & 25.7151 & 639 & -6.4 & $0.96\times0.92 \ (-45)$ & 40(8) & +7.1(0.1) & 1.8(0.1) & 1.0  \\
        \hline
          \multicolumn{5}{c|}{\multirow{2}{*}{NH$_3$}} &  \multicolumn{4}{c}{IRAS 4A1} \\
         \multicolumn{5}{c|}{} &  \multicolumn{4}{c}{$\rm 03^h29^m10\fs536$, $\rm 31^\circ13'31\farcs07$} \\
        \hline
        (3,3) & 23.8701 & 124 & -6.6 & $1.00\times0.95 \ (+6)$ & 121(16) & +6.4(0.1) & 2.7(0.2) & 1.3 \\
        (4,4) & 24.1394 & 201 & -6.5 & $0.99\times0.94 \ (-2)$ &  90(13) & +6.5(0.1) & 1.9(0.1) & 1.0 \\
        (5,5) & 24.5329 & 296 & -6.5 & $0.99\times0.95 \ (-3)$ &  87(12) & +6.4(0.1) & 2.5(0.1) & 1.0 \\
        (6,6) & 25.0560 & 409 & -6.5 & $0.97\times0.95 \ (-18)$ & 87(11) & +6.7(0.1) & 2.9(0.2) & 1.1 \\
        (7,7) & 25.7151 & 639 & -6.4 & $0.96\times0.92 \ (-45)$ & 46(9) & +6.4(0.1) & 1.9(0.2) & 1.1 \\
        \hline
    \end{tabular}
    \begin{minipage}{15cm}
        $^a$ Spectroscopic parameters are by \citet{xu_torsion_2008} from CDMS \citep[][]{muller_cologne_2005} for CH$_3$OH, and by \citet{yu_submillimeter-wave_2010} from JPL \citep[]{pickett_submillimeter_1998} for NH$_3$.\\
        $^b$ The spectral resolution is 1 km s$^{-1}$.\\
        %$^c$ The RMS is computed over each spectral window.
        %$^d$ Results of the Gaussian fit algorithm for CH$_3$OH, and the Hyper-fine fit algorithm for velocity and FWHM for NH$_3$ (see text).
    \end{minipage}
    %\end{threeparttable}
\end{table*}

\begin{figure*}
    \centering
    \includegraphics[scale=0.6]{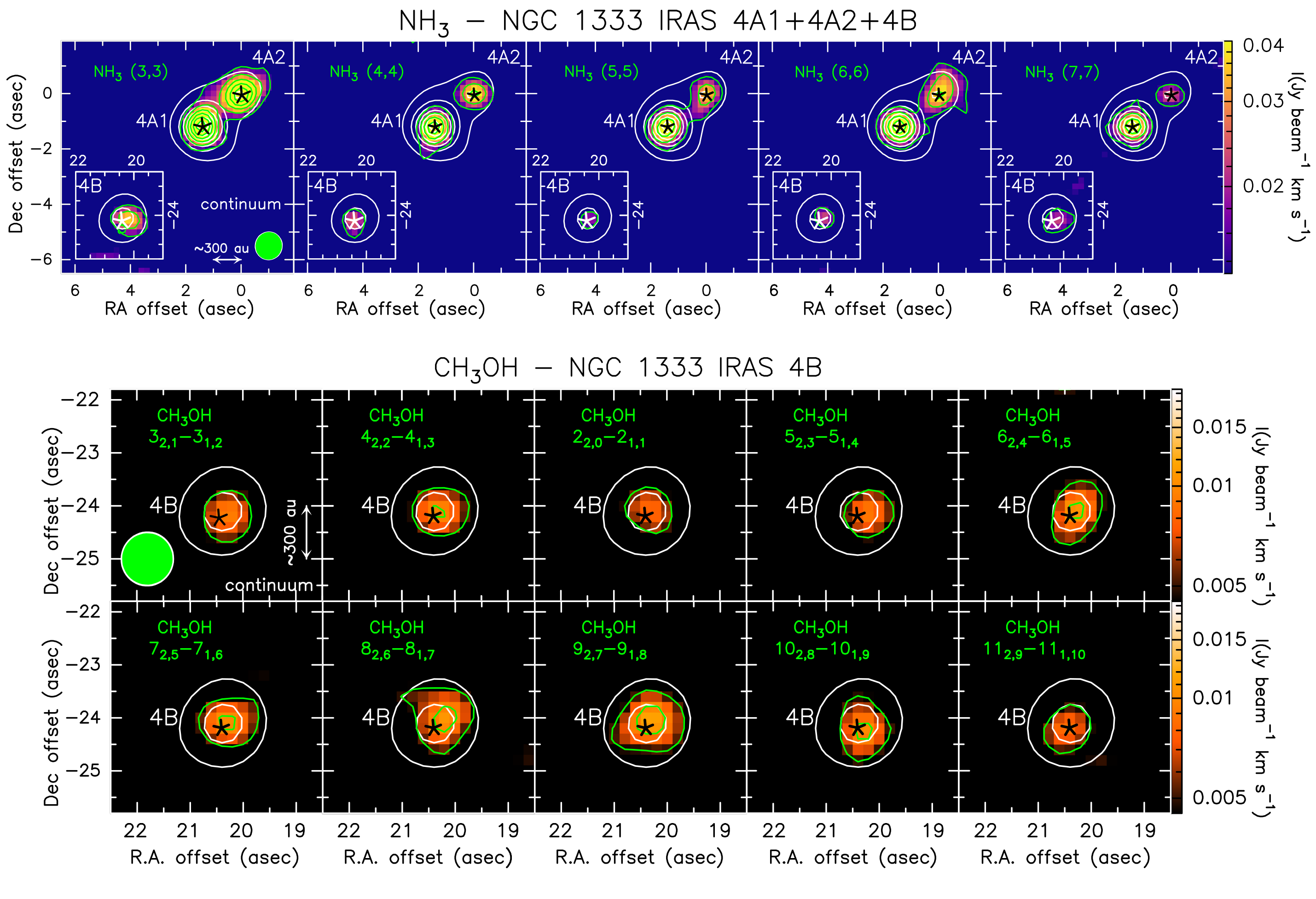}
    \caption{ \textit{Top panels}: NH$_3$ velocity-integrated maps  {(from $-$30 to $+$45 \kms with respect to the v$_{\rm sys}$, to include all the hyperfine components)} toward IRAS 4A1, 4A2, and 4B in color scale starting from 3$\sigma$, with green first contour and steps of 3$\sigma$ ($\sigma$=4.4 mJy beam$^{-1}$), overlapped with the continuum in white contours (from 50$\sigma$ with steps of 100$\sigma$). 
    \textit{Bottom panels:} CH$_3$OH velocity-integrated maps toward IRAS 4B (from $-$2 to $+$2 \kms with respect to the v$_{\rm sys}$) in color scale starting from 3$\sigma$, with green first contour and steps of 3$\sigma$ ($\sigma$=1.6 mJy beam$^{-1}$ \kms) overlapped with the continuum in white contours (from 50$\sigma$ with steps of 100$\sigma$). 
    The CH$_3$OH maps for 4A1 and 4A2 are reported in \citet{de_simone_hot_2020}.
    %The line transitions are reported in each panel.
    The stars mark the protostar positions. Synthesised beams are in the lower corners.
    }
    \label{fig:maps}
\end{figure*}

\begin{figure*}
    \centering
    \includegraphics[scale=0.5, angle=90]{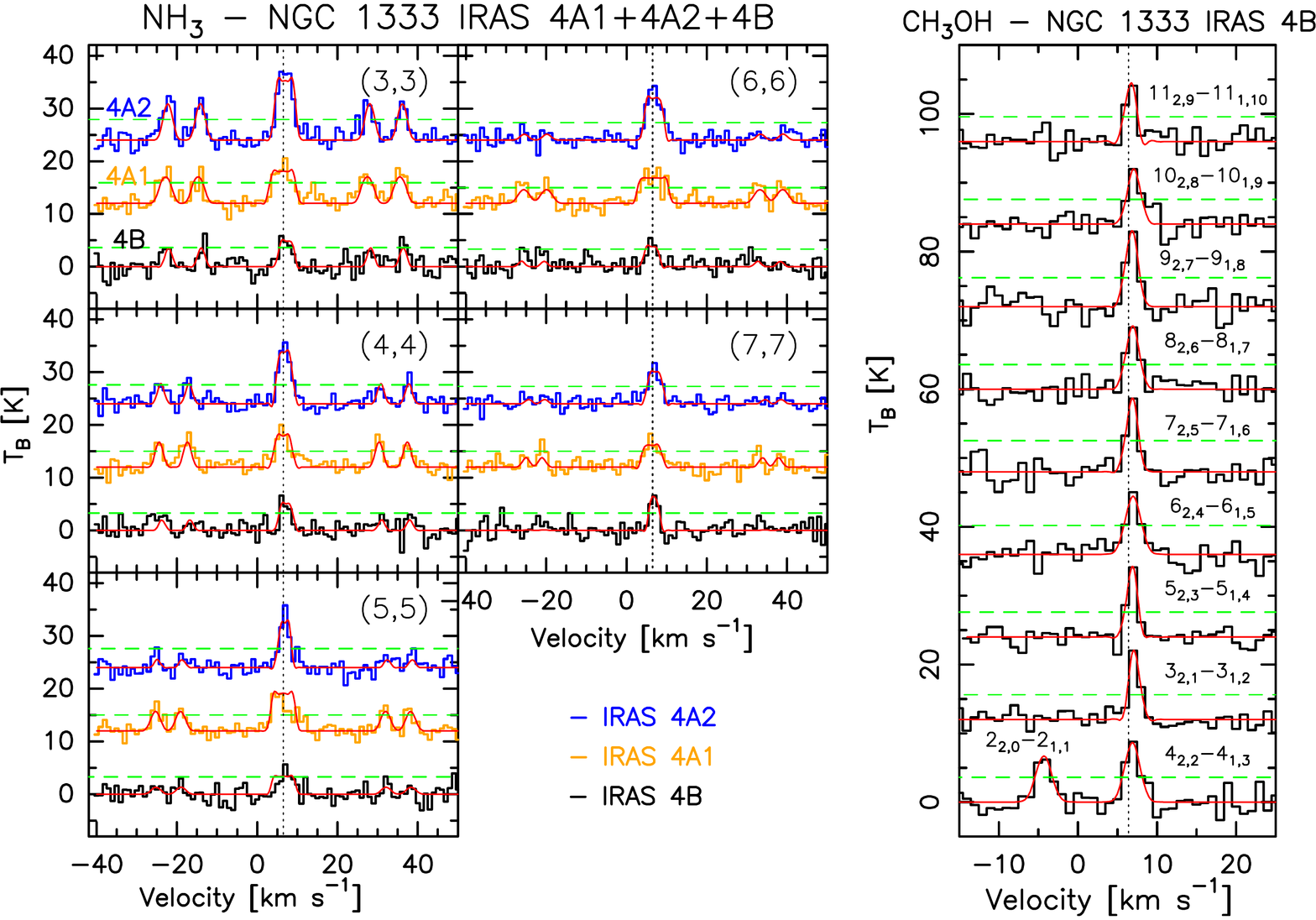}
    \caption{\textit{Left panels:} NH$_3$ lines (marked in each panel) detected toward the continuum peak of IRAS 4B (black), 4A1 (orange), 4A2 (blue). The red curves show the best hyperfine fits. \textit{Right panel:} CH$_3$OH lines (marked on each spectrum) detected in the VLA K-band toward 4B. The red curves show the best Gaussian fits. 
    In all panels, each spectrum is shifted by 12 K from the previous one, the vertical dotted black lines report the v$_{\rm sys}$ (6.7 \kms), and the horizontal green dashed lines show the 3$\sigma$ level.}
    \label{fig:spectra_tot}
\end{figure*}

%\subsection{Maps and Spectra}
Figure \ref{fig:maps} reports the NH$_3$ velocity-integrated map, for the targeted protostars (4A1, 4A2, and 4B), and the CH$_3$OH ones for 4B \citep[the 4A1 and 4A2 ones are in][]{de_simone_hot_2020}, overlapped with the continuum emission. All the targeted lines are detected with a S/N$\geq$5.
The molecular emission peaks at the protostellar continuum position (coordinates in Table \ref{tab:spectral_params&fit_res}) and it is not resolved at the current angular resolution. However, the emission of NH$_3$ \citep[and \meth;][]{de_simone_hot_2020} around 4A1 and 4A2 is well disentangled.
The spectra of the targeted lines extracted at the protostar continuum peak are shown in Figure \ref{fig:spectra_tot}. %\citep[the CH$_3$OH spectra for 4A1 and 4A2 are reported in][]{de_simone_hot_2020}.
 
We derived the velocity-integrated line intensities for each transition using a Gaussian fit for CH$_3$OH, and a Hyperfine fit for NH$_3$, having spectrally resolved its hyperfine structure.
The latter is assuming: i) the same excitation temperature and width for all the components, ii) a Gaussian distribution of velocity, and iii) not overlapping components.
Then, the NH$_3$ integrated emission is computed as the sum of the integrated area of the main and the satellite hyperfine components. 
The fit results 
%, namely the integrated emission ($\rm \int T_b dV$), the linewidth (FWHM), the peak velocities (V$\rm_{peak}$) and the RMS computed for each spectral window,
are reported in Table \ref{tab:spectral_params&fit_res}.
The velocity peaks are consistent with the systemic velocity of the cloud hosting the protostars ($\sim$ 6.7 \kms).

In summary, we detected and imaged multiple lines of ammonia (from (3,3) to (7,7)) and methanol towards the hot corinos, 4A1, 4A2 and 4B, at compact scale ($\geq 300$ au) around the central region.

%%%%%%%%%%%%%%%%%%%%%%%%%%%%%%%%%%%%%%%%%%%%%%%%%%%%%%%%%%%%%%%%%%%%%%%%%%%%%%%%%%%
\section{Radiative transfer and astrochemical modelling} \label{sec:modelling}

%\subsection{Column density estimation}
\subsection{Radiative transfer modelling}
Having detected several lines of NH$_3$ and \meth \ covering a large range of E$_{\rm up}$ (see Table \ref{tab:spectral_params&fit_res}) we performed a multi-line analysis to derive their abundance ratio. 
More specifically, we used a non-LTE analysis via our in-home Large Velocity Gradient (LVG) code \texttt{grelvg} \citep{ceccarelli_theoretical_2003} to predict the molecular line intensities that will be simultaneously fitted via comparison to the observed ones using a $\chi^2$ minimization.

The collisional coefficients of CH$_3$OH and NH$_3$ with para-H$_2$ are from the BASECOL database \citep{dubernet_basecol2012:_2013}. 
They are computed between 10 and 200 K by \citet{rabli_rotational_2010} for the first 256 levels of A- and E- \meth \ and by \citet{bouhafs_collisional_2017} for the lowest 17 and 34 levels of ortho- and para-NH$_3$, respectively.
We assumed a semi-infinite slab geometry to compute the line escape probability as a function of the line optical depth, the H$_2$ ortho-to-para ratio equal to 3, the CH$_3$OH A-type/E-type ratio equal to 1, and the NH$_3$ ortho-to-para ratio equal to 2. 
The latter will be discussed and justified a posteriori in  Section \ref{sec:discussion}.

\paragraph{Methodology}
A detailed description of the adopted methodology is in Section \ref{sec:App_method}, and a figurative scheme is shown in Figure \ref{fig:analysis_scheme}. Here we summarize the major steps.

We first performed the LVG analysis of the VLA methanol lines at 25 GHz in order to constrain the gas density and temperature, and to derive the \meth \ column density and emitting size \citep[the 4A1 and 4A2 ones have been derived in][]{de_simone_hot_2020}.
All \meth \ transitions were optically thick so that the derived column density was only a lower limit. 
To constrain the \meth \ column density we used the observations at mm wavelenghts \citep[from][for 4A2 and 4B, respectively,  {see Table \ref{tab:mm_lines}}]{taquet_constraining_2015,yang_perseus_2021} once corrected for the dust absorption factor (30\% at 143 GHz for 4A2, and of 50\% at 243 GHz for 4B; see Section \ref{sec:App_method}). 
 {For 4A1 there are no methanol mm-lines detected, so we report only a lower limit of its column density. }

Then, assuming that NH$_3$ traces the same gas as \meth\footnote{This assumption is verified a posteriori as described in Section \ref{sec:App_method}.}, namely assuming gas density and temperature ranges of \meth, we performed the LVG analysis of the NH$_3$ lines to derive the NH$_3$ column density and the emitting size. %retrieving the solutions for the same emitting size.

Finally, we computed the NH$_3$/\meth \ abundance ratio, using the column densities of NH$_3$ and \meth \ correspondent to the common derived source size.
 {The derived ratio in 4A1 is an upper limit due to the unconstrained methanol column density.}
The derived values are $\leq$ 0.5, 0.015--0.5, and 0.003--0.3 , for 4A1, 4A2 and 4B. 

The 1$\sigma$ confidence level ranges of the LVG fitting results are reported in Table \ref{tab:LVG_results+ratio}. 
As a result, both \meth \ and NH$_3$ are tracing compact ($<$100 au), dense ($>10^6$ cm$^{-3}$) and hot ($>100$ K) gas, fully consistent with the integrated emission maps (Figure \ref{fig:maps}) and the fact that we are observing the compact hot corino emission.

\begin{table}
    \centering
    \caption{Results of the non-LTE LVG analysis of CH$_3$OH and NH$_3$ toward IRAS 4A1, 4A2 and 4B using the \texttt{grelvg} code. The reported values are the 1$\sigma$ confidence level.
    }
    \label{tab:LVG_results+ratio}
    \hspace{-1.3cm}
    \resizebox{9.5cm}{!}{%
    \begin{tabular}{lc|cccc}
    \hline
         & & IRAS 4A1 & IRAS 4A2 & IRAS 4B  \\
         \hline
    n$_{ H_2}$ & [cm$^{-3}$] & $\geq$2$\times$10$^6$ & $\geq$7$\times$10$^6$ &  $\geq$2$\times$10$^6$ \\
    T$_{\rm kin}$ & [K] & 100-120 & 140-160  & 150-190\\
    N$_{\rm CH_3OH}$ & [cm$^{-2}$] & $\geq$ 10$^{19}$ & (0.6--4)$\times$10$^{19}$  & (1--8)$\times$10$^{19}$ \\
    N$_{\rm NH_3}$ & [cm$^{-2}$] & (1-5)$\times$10$^{18}$ & (0.6-3)$\times$10$^{18}$ &   (0.2-3)$\times$10$^{18}$\\
    size & [$"$] & 0.24-0.26 & 0.19-0.24 & 0.18-0.20 \\
    \hline
    ${\rm NH_3/CH_3OH}$ & - & $\leq$0.5 & 0.015--0.5 &  0.003--0.3\\
    \hline
   \end{tabular}
   }
   \begin{minipage}{8cm}
       %$^a$ The reported range for the source size is constrained from both CH$_3$OH and NH$_3$ analysis (see Section \ref{sec:App_method}). 
   \end{minipage}
\end{table}

\begin{figure*}
    \centering
    \includegraphics[scale=0.8]{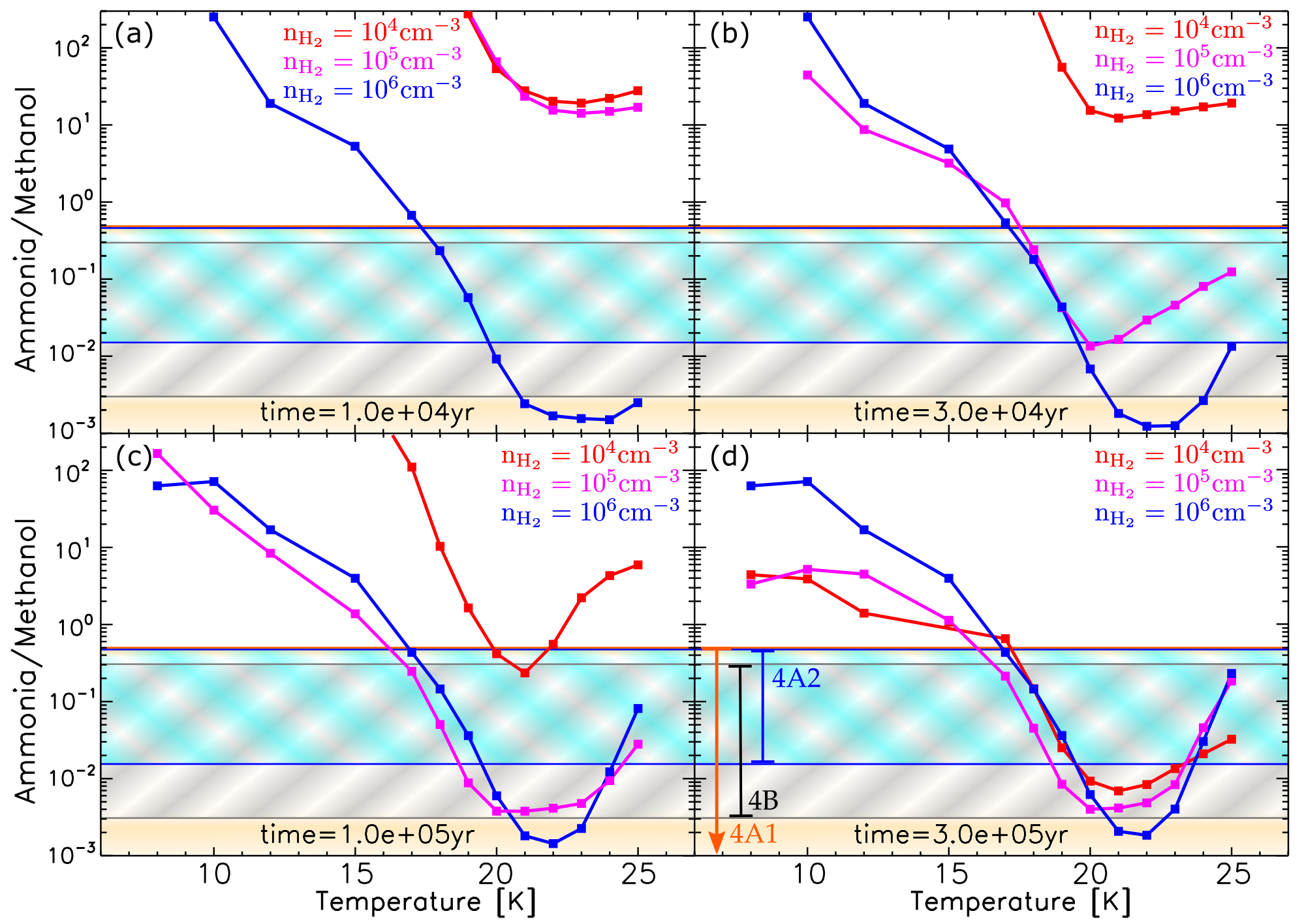}
    \caption{
    Theoretical predictions of NH$_3$/CH$_3$OH abundance ratio versus the pre-collapse dust temperature at different timescales ($\rm time=(0.1,0.3,1,3)10^5$ yr,  {in the (a), (b), (c), (d) panels respectively}) and H$_2$ density (n$_{H_2}=(0.1,1,10)10^5$ cm$^{-3}$, in  {red, magenta and blue respectively}), obtained with the \texttt{GRAINOBLE} code.  {Blue, grey and orange bands represent the NH$_3$/CH$_3$OH values derived from the LVG analysis in 4A2, 4B and 4A1, respectively.}
    }
    \label{fig:model}
\end{figure*}

%%%%%%%%%%%%%%%%%%%%%
\subsection{Astrochemical modelling} \label{subsec:astrochem-model}
 {The adopted model consists of two phases: (1) the mantle formation during the cold phase, and (2) the sublimation of the water-rich mantles when the dust temperature reaches the water sublimation temperature.
During the first phase, only a very small fraction ($\sim$1\%) of the frozen species is injected into the gas-phase by non-thermal desorption mechanisms  {\citep{minissale_dust_2016}}, so this does not impact the amount of frozen species in the grain mantles before the second step occurs.
During the second phase, the whole mantle sublimates so that the relative abundances observed in the gas-phase reflect the composition of the mantles. Our modeling focuses on the first phase, since the abundance on the grain mantles is the key point.}

\subsubsection{Model description}
We used the astrochemical model \texttt{GRAINOBLE} \citep{taquet_multilayer_2012, taquet_water_2013,ceccarelli_evolution_2018} to predict the evolution of the frozen NH$_3$/\meth \ abundance ratio as a function of the pre-collapse physical conditions (density, dust temperature, timescale). 

Briefly, it is a time-dependent 3-phase grain-gas chemistry code that computes the layered grain mantles structure.
The gas-phase reaction network is an updated version of the KIDA 2014 network \citep[\url{http://kida.obs.u-bordeaux1.fr};][]{wakelam_2014_2015}, with the reactions described in \citet{Tinacci_gretobape_2022}. 
The surface reactions are assumed to occur only in the last two formed mantle layers, the latter being in contact with the gas-phase.
In this work, we only considered the hydrogenation and oxidation of the species frozen on the grain mantle.
In general, when the required information is available, we used the Eckart formalism to describe the probability for a reaction with an activation barrier to occur \citep{taquet_water_2013}.
This is the case for the hydrogenation of CO into H$_2$CO and CH$_3$OH and the CO oxidation into CO$_2$.
In addition, hydrogenation of O, O$_2$ and O$_3$ leads to water, and of N to ammonia.
On the other hand, C hydrogenation cannot occur, as carbon atoms chemically bind with the water molecules of the ice \citep[e.g.,][]{shimonishi_adsorption_2018}. 
Methane is, therefore, formed by the hydrogenation of frozen CH.
The used binding energies are those reported in \cite{taquet_multilayer_2012}, updated using the ones computed by \citet{minissale_dust_2016,song_formation_2016,shimonishi_adsorption_2018,ferrero_binding_2020}, and  \citet{minissale_thermal_2022}. 
 {In particular, for CO, N, and N$_2$ we assumed 1750, 720, and 1300 K, respectively.}
The diffusion to binding energy is assumed to be 0.5, following recent experimental and theoretical results \citep[e.g.,][]{he_measurements_2018}.

We assumed that the H$_2$ number density n$_{H_2}$ of the molecular cloud is constant, and an average grain radius of 0.1 $\mu$m, typical of the Galactic ISM grains. %\citep[e.g.,][]{jones_evolution_2013}. 
The gas and dust are assumed to be thermally coupled.
 {To test that, we ran the model with the gas warmer than the dust. This scenario could be the result of an external shock, where there is a sudden increase in the gas temperature while the dust remains cool. Therefore, we assumed the gas temperature as high as 100 K \citep[typical temperature of shocked gas; e.g.,][]{codella_seeds_2017}.}
%However, we also ran cases with the gas warmer than the dust, with the gas temperature as high as 100 K ( {to simulate a sudden increase of the gas temperature due to, for example, a shock by an external event}) and 
The results are basically the same, so we will not discuss them in the following.

The initial elemental abundances were assumed to be the solar ones \citep{asplund_chemical_2009} depleted following \citet{jenkins_unified_2009}, where we assumed the most depleted cases: O/H=$2.8\times 10^{-4}$, C/H=$1.7\times 10^{-4}$ and N/H=$5.3\times 10^{-5}$.

We ran a grid of models with different H$_2$ density, $(0.1,1,10)\times 10^{5}$ cm$^{-3}$, and temperatures, from 8 to 25 K. 
Each model starts with all the elements in atomic form except hydrogen which is molecular and the chemical composition is left to evolve for 10$^7$ yr. 
Eventually, the formed mantle is constituted of approximately 100-160 layers, depending on the model parameters. 

\subsubsection{Model results} \label{subsec:mod-results}

We first verified that the steady state mantle composition of the model with  n$_{H_2}=10^4$ cm$^{-3}$ and T=10 K (a typical molecular cloud) is consistent with the observations of similar regions \citep{boogert_observations_2015}: H$_2$O/H$_2\sim1\times10^{-4}$, CO/H$_2\sim3\times10^{-5}$, CO$_2$/H$_2\sim3\times10^{-5}$, CH$_3$OH/H$_2\sim4\times10^{-5}$, NH$_3$/H$_2\sim2\times10^{-5}$.
This good agreement encourages the reliability of the predictions obtained with different densities and temperatures.

Figure \ref{fig:model} reports the theoretical predictions of the frozen NH$_3$/\meth \ as a function of the pre-collapse dust temperature at different timescales and H$_2$ density. 

\paragraph{Time dependence:} 
The NH$_3$/\meth \ abundance ratio decreases with time until it reaches a constant value after about (1, 3, 10)$\times 10^5$ yr, with H$_2$ density of (10, 1, 0.1)$\times10^5$ cm$^{-3}$, respectively. 
This is because methanol forms after ammonia.
First, when the N atoms land on the grain surfaces, they rapidly undergo hydrogenation, which is a barrierless process, while carbon is still atomic.
Then, the gaseous abundance of N drops because nitrogen goes into N$_2$, and CO forms. 
Finally, once the gaseous CO freezes out into the mantles, methanol is formed by CO hydrogenation. 

\paragraph{Temperature dependence:}
With the increase of the dust temperature, the residence time of N on the mantle decreases  {and, consequently, the ammonia abundance diminishes. 
On the contrary, the CH$_3$OH abundance on the mantle remains roughly constant until about 25--30 K, at which temperatures CO sublimates from the grain surfaces.
Therefore, the NH$_3$/\meth \ ratio decreases with increasing temperature}.
However, once the dust temperature becomes larger than the N$_2$ sublimation temperature ($\sim 20$ K), N$_2$ is released into the gas-phase, where it can react in the gas-phase to form gaseous NH$_3$ \citep[e.g.,][]{le_gal_interstellar_2014}. 
Since the dust temperature is still low, the gaseous NH$_3$ immediately depletes onto the grain surface (see Figure \ref{fig:scheme_nh3}) increasing again the NH$_3$/\meth \ ratio (see Figure \ref{fig:model}).

\paragraph{Density dependence:}
Increasing n$_{H_{2}}$, the curves shift towards shorter times, because of the higher accretion rate of the species in the mantle.
Indeed, the three curves coincide at time larger than $\sim 3\times 10^5$ yr (as shown in Figure \ref{fig:model}).

%%%%%%%%%%%%%%%%%%%%%%%%%%%%%%%%%%%%%%%%%%%%%%%%%%%%%%%%%%%%%%%%%%%%%%%%%%%%%%%%%%%
\section{Discussion} \label{sec:discussion}

%%% CC: penso che manchi una discussioncina di altri modelli e la dipendenza dei parameteri in input. Ne parliamo quando rientriamo next year.... --> discuss and cite aikawa2020
The first strong conclusion of the new observations is that the three protostars (4A1, 4A2, and 4B) possess a similar NH$_3$/\meth \ ratio (Figure \ref{fig:model}), which points to similar pre-collapse conditions.
This would be expected for IRAS 4A1 and 4A2, as they are coeval companions of a binary system.
However, it is not obvious for IRAS 4B, which is located $\sim$30$''$ ($\sim$ 9000 au) away from the binary system. 
In summary, the three protostars were born from pre-collapse material with similar physical conditions. 

Additionally, the comparison between the observations and the theoretical model predictions provides the following two strong constrains on the pre-collapse: 
(1) A collapse timescale less than $\sim10^4$ and $\sim10^5$ yr for a density of $10^6$ and $10^4$ cm$^{-3}$, respectively, cannot reproduce the observed NH$_3$/\meth \ ratio; 
(2) The pre-collapse dust temperature has to be larger than 17 K in all the three protostars. 
In other words, the grain mantles of the three protostars was formed during a period not smaller than $\sim10^4$--10$^5$ yr, depending on the cloud density.

Most importantly, the dust was relatively warm, about 17 K.
We emphasize that this dust is the one corresponding to the inner $<100$ au of the hypothetical condensations from which the three protostars were born.
If we consider the typical temperature of a prestellar core at this scale, we would expect much lower temperatures, around 7 K \citep[as for the prototypical pre-stellar core L1544;][]{crapsi_observing_2007}. Therefore, the results of 17 K is apparently puzzling.
On the other hand, the large scale maps of Herschel-Planck \citep{zari2016} show that the average dust temperature of the south part of NGC 1333 is around 17 K, with the denser parts at 14--15 K \citep{zhang_herschel_2022}.
Probably the only way to reconcile this ensemble of information is that the three protostars actually did not have the usual dense and cold pre-collapse phase period, as their mantles were mostly built during a relatively warm phase (dust temperature $\sim 17$ K) which is characteristic of the less dense cloud material in NGC 1333 south.
In other words, something must have happened that suddenly compressed the gas and triggered a fast collapse and the protostars' formation.

It is well known that the NGC 1333 region is heavily shaped by external triggers. 
In particular, it has been suggested that the filament where the three protostars lie could have been shaped by a colliding turbulent cell that would have triggered the birth of the protostars \citep{dhabal_connecting_2019}. 
The recent detection of a train of finger-shaped shocked SiO-emitting gas around IRAS 4A supports that an expanding bubble clashed against the southern part of NGC 1333 \citep{de_simone_bubble_2021}.
Our new analysis adds a new element to the story: the clash has brutally started the collapse in a region where otherwise no pre-collapse cores existed.

Finally, the derived dust temperature at the time of the mantle formation, around 17 K, justify a posteriori our choice of a NH$_3$ ortho-to-para ratio equal to 2.
This value corresponds to the thermal equilibrium at 15 K \citep[e.g.][]{faure2013}, and applies if ammonia was mostly formed on the icy grain surfaces.
We emphasise that the results of the analysis would not significantly change if a ratio equal to 1 (appropriate for larger temperatures) is adopted.

\section{Conclusions}
We observed NH$_3$ and \meth \ lines at cm wavelenghts with the VLA, toward the NGC 1333 IRAS 4A1, 4A2, and 4B protostars, finding that they are tracing the compact ($<100$ au) hot corino region. Using a non-LTE analysis we derived similar NH$_3$/\meth \ abundance ratios for all three protostars ($\leq$ 0.5 for IRAS 4A1, 0.015--0.5 for IRAS 4A2, and 0.003-0.3 for IRAS 4B). This means that they were born from pre-collapse material with similar physical conditions. 

Comparing the observed ratio with astrochemical models we constrained the pre-collapse conditions, finding that the dust was particularly warm ($\geq$17K). In other words, the protostellar ice mantles were mostly formed during a warm phase that it is typical of the less dense material of NGC 1333 southern region. 
We conclude that the collapse could have been brutally started by the clash of an external bubble with NGC 1333 in a warm region where no pre-collapse core existed. 
 
%This work shows how powerful is this method in retrieving the ice mantle history of Solar-type protostars without being biased by the dust opacity, and consequently in constraining the dynamical history of the hosting star forming region. 

%%Indeed, with interferometric observations at centimeter wavelengths it is possible to get rid of the dust contribution (that can be very large in the embedded protostellar phase) and to observe simultaneously, at planet formation scales, the two major grain mantle components NH$_3$ and CH$_3$OH, whose relative abundance, being grain surface species, will mainly depend on the gas conditions during their formation in the prestellar phase.
Finally, these results  {advance} the study of the chemical and dynamical history of protostars and open the way to future projects with the upcoming centimeter facilities such as ngVLA\footnote{\url{https://ngvla.nrao.edu/}} and SKA\footnote{\url{https://www.skatelescope.org/}}.
 {Additionally, the synergy between the upcoming centimeter facilities and the infrared ones (e.g., JWST \footnote{\url{https://www.jwst.nasa.gov/}} and ELT \footnote{\url{https://elt.eso.org/}}, which will provide the ice mantle composition along the line of sight of protostars and protoplanetary disks) will be crucial for characterizing the chemical and physical evolution of the early stages of planetary system formation.}

%%%%%%%%%%%%%%%%%%%%%%%%%%%%%%%%%%%%%%%%%%%%%%%%%%%%%%%%%%%%%%%%%%%%%%%%%%%%%%%%%%%%%%%%%%%%%%%%%%%%%%%%%%%%%%%%%%%%%%%%%%%%%%
%%%%%%%%%%%%%%%%%%%%% STOP COUNT HERE %%%%%%%%%%%%%%%%%%%%%%%%%
%%%%%%%%%%%%%%%%%%%%%%%%%%%%%%%%%%%%%%%%%%%%%%%%%%%%%%%%%%%%%%%%%%%%%%%%%%%%%%%%%%%%%%%%%%%%%%%%%%%%%%%%%%%%%%%%%%%%%%%%%%%%%%
\acknowledgements
The authors thank the anonymous referee for the constructive comments, which helped to improve the quality of the paper.
This work has received funding from the European Research Council (ERC) under the European Union's Horizon 2020 research and innovation programme, for the Project ``The Dawn of Organic Chemistry'' (DOC), grant agreement No 741002.
H.B.L. is supported by the Ministry of Science and Technology (MoST) of Taiwan (Grant Nos. 108-2112-M-001-002-MY3 and 110-2112-M-001-069-).
% {BE CAREFUL: THERE ARE TOO MANY REFERENCES (now 61, MAX 50)}

\appendix
\counterwithin{figure}{section}

%\section{NH$_3$ and CH$_3$OH spectra} \label{app:spectra}
%Figure \ref{fig:spectra_tot} shows the spectra extracted toward the continuum peak of the protostars. In particular, the NH$_3$ lines for 4A1, 4A2, and 4B, and the \meth \ lines for 4B. The \meth \ spectra for 4A1 and 4A2 are reported in \citet{de_simone_hot_2020}.

\section{NH$_3$ and \meth \ line analysis} \label{app:analysis} \label{sec:App_method}
\begin{figure}
    \centering
    \includegraphics[scale=0.5]{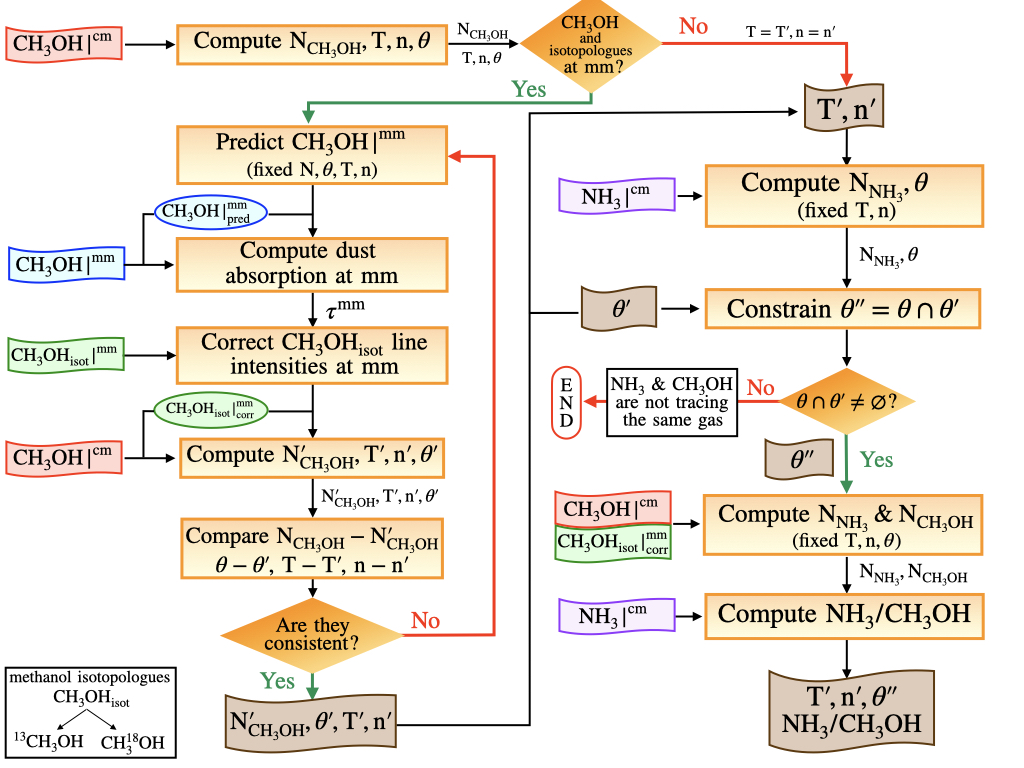}
    \caption{Scheme of the \meth \ and NH$_3$ line analysis performed to compute the NH$_3$/\meth \ abundance ratio in the three protostars. 
    The \meth \ lines at cm wavelengths (red flag) are the ones observed at 25 GHz with the VLA (this work). 
    The \meth \ lines at mm wavelengths (blue flag) and its isotopologues \meth$_{\rm isot}$ (green flag) are from \citet[][]{taquet_constraining_2015} for 4A2 using \meth$_{\rm isot}=^{13}$\meth, and from \citet[][]{yang_perseus_2021} for 4B using \meth$_{\rm isot}=$CH$_3^{18}$OH.
    For 4A1 we do not have mm methanol detections.
    All the computations and predictions of column density N, temperature T, density n and size $\theta$ are performed with a non-LTE LVG method.
    }
    \label{fig:analysis_scheme}
\end{figure}

\begin{table}[]
\caption{Spectral parameters, synthesized beams and integrated flux for CH$_3$OH, $^{13}$CH$_3$OH, and CH$_3^{18}$OH transitions at mm wavelengths toward IRAS 4A2 and IRAS 4B from \citet{taquet_constraining_2015} and \citet{yang_perseus_2021}.}\label{tab:mm_lines}
    \centering
    \begin{tabular}{l|ccccc}
    \hline
    \hline
    \multirow{2}{*}{Transition } & \multirow{2}{*}{Frequency$^{(a)}$} & \multirow{2}{*}{E$_{\rm up}^{(a)}$}& \multirow{2}{*}{logA$_{\rm ij}^{(a)}$} & Synthesized Beam &  \\
        %& \multicolumn{4}{c|}{IRAS 4A2} & \multicolumn{4}{c}{IRAS 4A1} \\
    &   &   &   & maj $\times$ min (PA)  &   $\rm \int T_BdV$  \\
    & [GHz] & [K] &  & [$''\times ''$ ($^\circ$)] & [K km s$^{-1}$]  \\
    \hline
    \multicolumn{6}{c}{IRAS 4A2 \citep{taquet_constraining_2015}} \\
    \hline
     & \multicolumn{5}{c}{CH$_3$OH} \\
    \hline
    3(1,3)-2(1,2) A & 143.8658 & 28 & -5.0 & 2.2$\times$1.8(+25) & 6.5(1.5) \\
    7(3,5)-8(2,7) E & 143.1695 & 113 & -5.4 & 2.3$\times$1.8(+26) & 3.6(1.2) \\
    \hline
    & \multicolumn{5}{c}{${^{13}}$CH$_3$OH} \\
    \hline
    3(0,3)-2(0,2) A & 141.6037 & 14 & -4.9 & 2.1$\times$1.7(+26) & 1.7(0.5) \\
    3(1,2)-2(1,1) A & 142.8077 & 28 & 4.9 & 2.1$\times$1.7(+26) & 1.2(0.3) \\
    6(2,5)7(1,6) A & 142.8967 & 85 & -5.3 & 2.1$\times$1.7(+26) & 1.5(0.5) \\
    7(0,7)-6(1,5) E & 163.8729 & 76 & -5.0 & 2.4$\times$1.8(+114) & 1.4(0.4) \\
    6(2,4)-7(1,7) A & 165.2805 & 85 & -5.1 & 2.4$\times$1.8(+114) & 0.8(0.2) \\
    13(1,12)-12(2,11) A & 165.2805 & 222 & -5.0 & 2.4$\times$1.8(+114) & 1.3(0.4) \\
    2(1,1)-2(0,2) E & 165.5756 & 28 & -4.6 & 2.4$\times$1.8(+114) & 1.8(0.4) \\
    3(1,2)-3(0,3) E & 165.6094 & 35 & -4.6 & 2.4$\times$1.8(+114) & 2.0(0.4) \\
    4(1,3)-4(0,4) E & 165.6909 & 44 & -4.6 & 2.4$\times$1.8(+114) & 1.8(0.4) \\
    5(1,4)-5(0,5) E & 165.3693 & 55 & -4.6 & 2.4$\times$1.8(+114) & 1.8(0.4) \\
    6(1,5)-6(0,6) E & 166.1287 & 69 & -4.6 & 2.4$\times$1.8(+114) & 1.8(0.4) \\
    7(1,6)-7(0,7) E & 166.5694 & 85 & -4.6 & 2.4$\times$1.8(+114) & 2.4(0.5) \\
    \hline
    \hline
    \multicolumn{6}{c}{IRAS 4B \citep{yang_perseus_2021}}\\
    \hline
    & \multicolumn{5}{c}{CH$_3$OH} \\
    \hline
    5(1,4)-4(1,3) A & 243.91579 & 50 & -4.2 & 0.6$\times$0.4(0) & 21(2) \\
    \hline 
     & \multicolumn{5}{c}{CH$_3^{18}$OH} \\
    \hline
    11(2,10)-10(3,7) A & 246.2566 & 184 & -4.6 & 0.6$\times$0.4(0) & 6.0(1.6) \\
    \hline
    \end{tabular}
\end{table}
Figure \ref{fig:analysis_scheme} shows the scheme of the \meth \ and NH$_3$ line analysis performed to compute the NH$_3$/\meth \ abundance ratio in the three protostars. All the computations and predictions of column density, temperature, density and size are performed with a non-LTE LVG method \citep{ceccarelli_theoretical_2003} with the following strategy:

\paragraph{Methanol line analysis}
The methanol line analysis for 4A1 and 4A2 is reported in \citet{de_simone_hot_2020}. We carried out a similar analysis for 4B. We ran a large grid of models ($\sim$ 70000) covering the frequency of the observed \meth \ lines, a total (A-type plus E-type) column density N$_{\rm CH_3OH}$ from $2\times10^{16}$ to $16\times10^{19}$ cm$^{-2}$, a gas density n$_{H_2}$ from  $10^6$ to $10^9$ cm$^{-3}$, both sampled in logarithmic scale, and a temperature T from 40 to 200 K, sampled in linear scale. 
We simultaneously fit the measured \meth \ line intensities, for 4B, via comparison with those simulated by the LVG model, leaving N$_{\rm CH_3OH}$, n$_{H_2}$, T and the emitting size $\theta$ as free parameters. 
Following the observations, we assumed a linewidth equal to 2 \kms \ and we included the calibration uncertainty (15\%) in the observed intensities. 
Solutions with N$_{\rm CH_3OH}\geq10^{18}$ cm$^{-2}$ are within 1$\sigma$ of confidence level, emitted by a source of $0\farcs18-0\farcs40$.  The  {reduced $\chi^2$ ($\chi_R^2$, defined as the $\chi^2$ per degree of freedom)} decreases, increasing the \meth \ column density, until a constant value ($\chi_R^2\sim$0.7), as all the observed lines  become optically thick ($\tau=1.5-4$) and, consequently, the emission is that of a black body. 

\paragraph{Constrain methanol column density}
For all the three sources, all the methanol lines observed at 25 GHz are optically thick so that we could derive only a lower limit for the methanol column density. 
To constrain the methanol column density we used its isotopologues observed at mm wavelengths, once corrected for the absorption factor due to the dust. 

While for 4A1 there are no methanol lines detected at mm wavelengths, for 4A2 there are several detected lines of CH$_3$OH and $^{13}$CH$_3$OH at 143 GHz by \citet{taquet_constraining_2015}, and for 4B there are three detected CH$_3$OH and one CH$_3^{18}$OH line at 243 GHz by \citet{yang_perseus_2021}\footnote{Please note that we could not use the lines at 243 GHz by \citet{yang_perseus_2021} for 4A2 as the methanol lines where contaminated by the outflow and we could not derive a reliable dust absorption factor.}  {(see Table \ref{tab:mm_lines})}.
 {Note that the angular resolution of the mm observations are slightly different from our cm observations. We took that into account in the computation of the filling factor when performing the LVG analysis. }

As explained in \citet{de_simone_hot_2020}, we retrieved the correction factor due to the dust contribute as follows: 
i) we predicted the methanol intensities at millimeter wavelengths using the gas conditions derived at cm wavelenghts with the VLA methanol lines; ii) we compared the predicted methanol millimeter line intensities with the observed ones; iii) we derived the dust optical depth at millimeter wavelength using Equation 1 in \citet{de_simone_hot_2020}, and consequently the dust absorption factor.  
%With the methanol lines, we retrieved the (foreground) dust absorption factor as explained in \citet{de_simone_hot_2020},
We found an absorption of 30\% at 143 GHz for 4A2, and of 50\% at 243 GHz for 4B. 

We then corrected the integrated line intensities of the detected methanol isotopologues ($^{13}$CH$_3$OH lines at 143 GHz for 4A2, and CH$_3^{18}$OH line at 243 GHz for 4B) by the derived dust absorption factor, and we run again the LVG code including the isotopologues together with the methanol lines at 25 GHz. We assumed a $^{12}$C$/^{13}$C of $\sim$70 \citep{milam_isotope_2005} and $^{16}$O$/^{18}$O$\sim$560 \citep{wilson_abundances_1994}. 
Since the $^{13}$CH$_3$OH and CH$_3^{18}$OH lines are optically thin, we could constrain the column density of methanol for 4A2 and 4B, finding 0.6--4$\times10^{19}$ cm$^{-2}$ and 1--8$\times10^{19}$ cm$^{-2}$, respectively. 
The gas density and temperature, the \meth \ column density and emitting size for IRAS 4A1, 4A2 and 4B are reported in Table \ref{tab:LVG_results+ratio}. 

\paragraph{Ammonia line analysis} 
We then considered the source size, the gas density and temperature ranges derived for methanol and ran a grid of models ($\sim$ 5000) with these parameters to fit the observed ammonia lines for the three protostars, leaving the ammonia column density as free parameter. 

Following the observations we assumed a linewidth of 12, 10, 9 \kms, for 4A1, 4A2, and 4B, respectively, computed as the one derived from the Hyperfine fitting (Table \ref{tab:spectral_params&fit_res}) multiplied by the number of components (main plus satellites). 
This is because we considered, in the LVG analysis, only the rotational level of NH$_3$ for which we have the collisional coefficients. 
The best fit is obtained for N$_{\rm NH_3}= 2\times10^{18}$ cm$^{-2}$ with $\chi_R^2$=0.5 for 4A1, $2\times10^{18}$ cm$^{-2}$ with $\chi_R^2$=1.1 for 4A2, and $6\times10^{17}$ cm$^{-2}$ with $\chi_R^2$=0.8 for 4B. %, emitted by source of $0\farcs22-0\farcs30$, $0\farcs20-0\farcs24$, $0\farcs20-0\farcs26$, respectively . 
The 1$\sigma$ confidence level range is reported in Table \ref{tab:LVG_results+ratio}. 
The NH$_3$ (3,3) transition is often quite optically thick ($\sim$20), while the others have opacity between 0.4 and 7. The (7,7) transition is always optically thin ($\leq$1), so it allowed us to constrain the NH$_3$ column density. 

Please note that we also run the LVG model on the NH$_3$ alone leaving column density, temperature, density and source size as free parameters. 
{We compared the ammonia parameter space with that derived from the methanol analysis: the two parameter spaces overlap with the one of methanol being better constrained. Therefore, we proceeded as described above.}
%and we found values consistent with the one derived with the analysis described above but with larger uncertainties. Indeed, due to the lower number of NH$_3$ lines with respect to CH$_3$OH, the derived ranges were not well constrained, therefore, we decided to proceed as described above. 

\paragraph{Abundance ratios}
Finally, we computed the NH$_3$/\meth \ abundance ratio, using the column densities of NH$_3$ and \meth \ corresponding to the common source size derived from the LVG analysis described above. 

The assumption that the two species are tracing the same gas is supported by the emission maps (Figure \ref{fig:maps}) and by the fact that a common source size has consistently been derived. 
{Additionally, there is no theoretical reason for why they would trace different gas on these size scales, as they are released together into the gas phase once the ice mantles are sublimated. }
The obtained NH$_3$/\meth \ values are $\leq$ 0.5, 0.015--0.5, and 0.003-0.3 , for 4A1, 4A2 and 4B (Table \ref{tab:LVG_results+ratio}). 
{Note that for 4A1 we could derive only an upper limit for the ratio, since we could not constrain the methanol column density for the lack of methanol emission at millimeter wavelengths. }

\section{NH$_3$ Formation}
\begin{figure}
    \centering
    \includegraphics[width=0.7\textwidth]{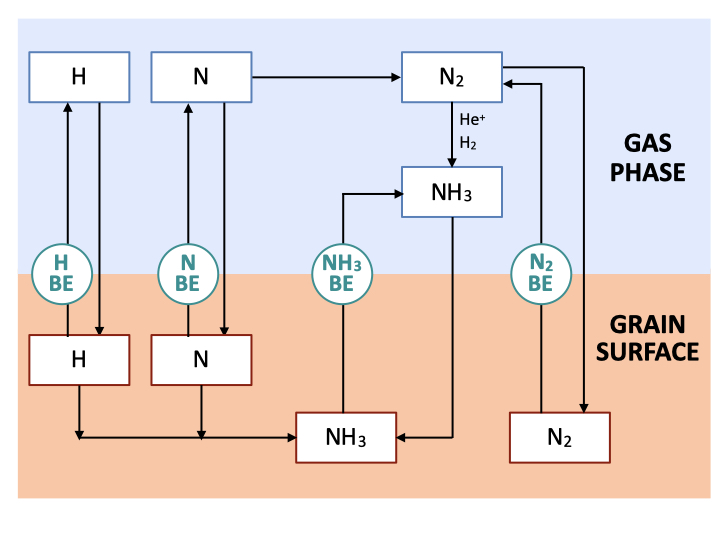}
    \caption{Scheme of the interplay between gas-phase and grain surface chemistry for the NH$_3$ formation \citep[{adapted} from ][]{Tinacci_theoretical_2022}. The release of H, N, N$_2$ and NH$_3$ into the gas-phase is regulated by their respective binding energies.  {BE stands for Binding Energy}.
    }
    \label{fig:scheme_nh3}
\end{figure}

Figure \ref{fig:scheme_nh3} \citep[from][]{Tinacci_theoretical_2022} shows the interplay between the gas-phase and the grain surface chemistry for the NH$_3$ formation. 
The major NH$_3$ formation path is through the hydrogenation of frozen N on the grain surfaces \citep{jonusas2020}, as it is a fast and barrierless process. 
When the dust temperature is high enough to release N$_2$ into the gas-phase for thermal desorption, the gas-phase pathway to form NH$_3$ takes place \citep{le_gal_interstellar_2014}. However, the temperature is still low  {(around 20--25 K)} for NH$_3$ to remain into the gas-phase, therefore it freezes out onto the grain. 
Once on the grain surface, NH$_3$ can be thermally desorbed or injected into the gas-phase via the so-called Chemical Desorption (CD). While CD injects a small fraction  {\citep[$\leq1$\%][]{minissale_dust_2016}} of the NH$_3$ into the gas-phase, the thermal desorption, governed by the NH$_3$ Binding Energy (BE), involves the whole frozen NH$_3$ .
{In our case, NH$_3$ will be released into the gas phase through the sublimation of the icy mantles when the dust temperature reaches the water sublimation temperature (above 100 K). }

\bibliographystyle{aasjournal}
\bibliography{IRAS4}{}

\end{document}